\documentclass[conference]{IEEEtran}
\IEEEoverridecommandlockouts
\usepackage{cite}
\usepackage{amsmath,amssymb,amsfonts}
\usepackage{algorithmic}
\usepackage{graphicx}
\usepackage{textcomp}
\usepackage{float}
\usepackage{url}
\usepackage{booktabs}
\usepackage{multirow}
\usepackage[table,xcdraw]{xcolor}
\usepackage[section]{placeins}
\usepackage{multicol}

\def\BibTeX{{\rm B\kern-.05em{\sc i\kern-.025em b}\kern-.08em
    T\kern-.1667em\lower.7ex\hbox{E}\kern-.125emX}}

\begin{document}
\onecolumn

\
\vfil
\noindent Preprint Notice:
\newline
© 2019 IEEE. Personal use of this material is permitted. Permission from IEEE
must be obtained for all other uses, in any current or future media, including
reprinting/republishing this material for advertising or promotional purposes,
creating new collective works, for resale or redistribution to servers or lists, or
reuse of any copyrighted component of this work in other works.
\vfil

\twocolumn
\newpage
% \onecolumn % delete to make it right again

\title{3D Deformable Convolutions for MRI classification}

\author{\IEEEauthorblockN{1\textsuperscript{st} Marina Pominova*}
\IEEEauthorblockA{\textit{Skolkovo Institute}\\
\textit{of Science and Technology}\\
Moscow, Russia \\
marina.pominova@skoltech.ru}\\

\IEEEauthorblockN{5\textsuperscript{th} Alexander Bernstein}
\IEEEauthorblockA{\textit{Skolkovo Institute}\\
\textit{of Science and Technology}\\
Moscow, Russia \\
a.bernstein@skoltech.ru}

\and

\IEEEauthorblockN{2\textsuperscript{nd} Ekaterina Kondrateva*}
\IEEEauthorblockA{\textit{Skolkovo Institute}\\
\textit{of Science and Technology}\\
Moscow, Russia \\
ekaterina.kondrateva@skoltech.ru}\\

\IEEEauthorblockN{6\textsuperscript{th} Evgeny Burnaev}
\IEEEauthorblockA{\textit{Skolkovo Institute}\\
\textit{of Science and Technology}\\
Moscow, Russia \\
e.burnaev@skoltech.ru}

\and

\IEEEauthorblockN{3\textsuperscript{rd} Maksim Sharaev}
\IEEEauthorblockA{\textit{Skolkovo Institute}\\
\textit{of Science and Technology}\\
Moscow, Russia \\
m.sharaev@skoltech.ru}

\and

\IEEEauthorblockN{4\textsuperscript{th} Sergey Pavlov}
\IEEEauthorblockA{\textit{Moscow Institute}\\
\textit{of Physics and Technology}\\
Moscow, Russia \\
sergey.pavlov@phystech.edu}

}
\maketitle

% \begin{multicols}{2} % delete to make it right again

\begin{abstract}
Deep learning convolutional neural networks have proved to be a powerful tool for MRI analysis. In current work, we explore the potential of the deformable convolution deep neural network layers for MRI data classification.  
We propose new 3D deformable convolutions (d-convolutions), implement them in VoxResNet architecture and apply for structural MRI data classification. 
We show that 3D d-convolutions outperform standard ones and are effective for unprocessed 3D MR images being robust to particular geometrical properties of the data. Firstly proposed \textit{dVoxResNet} architecture exhibits high potential for the use in MRI data classification.

\end{abstract}


\section{Introduction}
% problem statement 
There is a need for accurate prediction and diagnostics in neurology and psychiatry. MRI is considered as one of the most powerful diagnostic 
instruments applicable for multiple examinations both in adults and children, see \cite{bernhardt2015magnetic, hosseini2016alzheimer, Pominova2019,DepressionAWE2018}.

% Рments, applicaеc toolвью методов (ссылки на наши статьи старые) #Максим
% 3.  Какие есть проблемы #Катя
{There is a number of successful applications of convolutional neural networks (CNN) with different architectures for segmentation of  MRI data. Many of these solutions are based on adapting existing approaches to analyzing 2D images for processing of three-dimensional data. 

For example, for brain segmentation, an architecture similar to ResNet \cite{b1} was proposed, which expands the possibilities of deep residual learning for processing volumetric MRI data using 3D filters in convolutional layers. The model, called VoxResNet \cite{chen2018voxresnet}, consists of volumetric residual blocks (VoxRes blocks), containing convolutional layers as well as several deconvolutional layers. The authors demonstrated the potential of ResNet-like volumetric architectures, achieving better results than many modern methods of MRI image segmentation \cite{milletari2016v}. Convolutional neural networks also showed good classification results in problems associated with neuropsychiatric diseases such as Alzheimer's disease \cite{hosseini2016alzheimer}. Results were more accurate or comparable to earlier feature based approaches that use extracted morphometrical lower dimensional brain characteristics \cite{milletari2017hough, zou20173d, farooq2017deep,chen2018mri }.

Thus, convolutional networks can be applied directly to the raw neuroimaging data without loss of model performance and over-fitting, which allows skipping the feature extraction step.}

However, there is a problem: the traditional convolutional networks are very sensitive to image size, scale and spatial orientation thus require thorough pre-processing specific to a clinical application.
Because different locations in the input feature maps may correspond to objects with different scales or deformation, adaptive determination of scales or receptive field sizes is desirable for certain tasks, particularly in neuroimaging.

It is known that CNNs are inherently limited to model large unpredictable transformations because of the fixed geometric structure of the sampling matrix, and restricted receptive fields.

In the present study we carried out an extensive experimental evaluation of deep CNNs both with traditional and deformable convolutions for bipolar disorder and schizophrenia diagnostics based on structural MRI data. The article has the following structure: in Section 2 we review current deep learning approaches used for MRI data analysis and their drawbacks as well as a possible solution --- deformable convolutions. Here we also present  the training datasets and the classification task. We describe obtained results in Section 3, provide discussion of the network performance in Section 4 and draw conclusions in Section 5.

\section{Materials and Methods}
\label{sec:review}

\subsection{MRI data characteristics}

In magnetic resonance imaging (MRI), strong magnetic fields are used to create images of biological tissues and physiological changes in them. MRI data, i.e. 3D brain images, are often noisy and high dimensional.

MR noise is associated with the scanning procedure (low-level hardware artefacts such as magnetic field inhomogeneity, radiofrequency noise, surface coil artefacts and others) and signal processing (chemical shift, partial volume, etc.); as well as with the scanned patient (physiological noise such as blood flow, movements, etc.) \cite{Erasmus2004}.  

In addition to MRI data cleaning problem, there is another common challenge of the brain imaging analysis related to big data dimensionality, which mostly depends on resolution parameters of the scanner inductive detection coil. For instance, standard voxel sizes are within $0.5$--$2$\,mm$^3$ in case of structural imaging (resulting in $10^7$ voxels for the whole brain volume). Thus an MRI image, composed of a huge number of small-sized voxels, has higher spatial resolution and, hence, high dimensionality.

A canonical approach to structural MRI data analysis is morphometry features extraction: brain structure is first segmented into anatomical regions or tissue types, and then different characteristics like region volumes, tissue thickness and many others are calculated. These features are then used in regression or classification algorithms.

\begin{figure}[ht]
\begin{center}
\centerline{\includegraphics[width=\columnwidth]{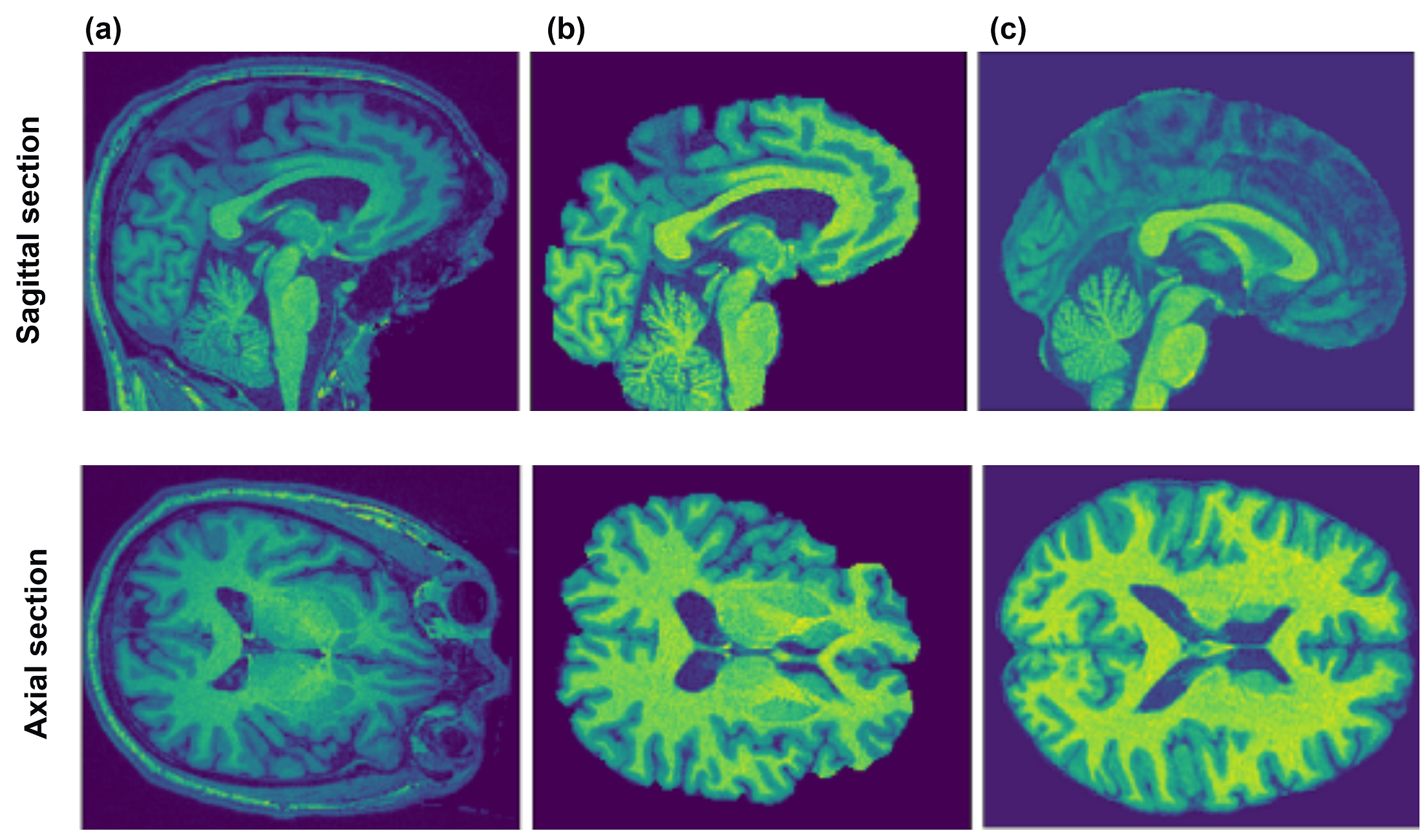}}
\caption{Sagittal and  Axial sections view of structural MRI images on different preprocessing stages: (a) imagery without preprocessing after anonymization, (b) skull-stripped imagery, (c) skull-stripped and MNI normalized images. }
\label{fig:brain_preprocessing}
\end{center}
\vskip -0.2in
\end{figure}

\subsection{Deformable Convolutions}
Even after the same preprocessing, MRI data from different scanners can vary significantly in size and aspect ratio of brain images.

There are two natural ways to make the model more stable with respect to such changes in spatial characteristics. First, we can augment the training data by scaling images by random values along different axes.
% Robust representations can be learned from the data, but usually at the cost of expensive training and complex model parameters.
Second, we can make the network itself invariant to transformations of the input image, introducing certain modification to its architecture. The latter can be implemented using deformable convolutions \cite{dai2017deformable}.

\begin{figure}[ht]
\vskip 0.2in
\begin{center}
\centerline{\includegraphics[width=\columnwidth]{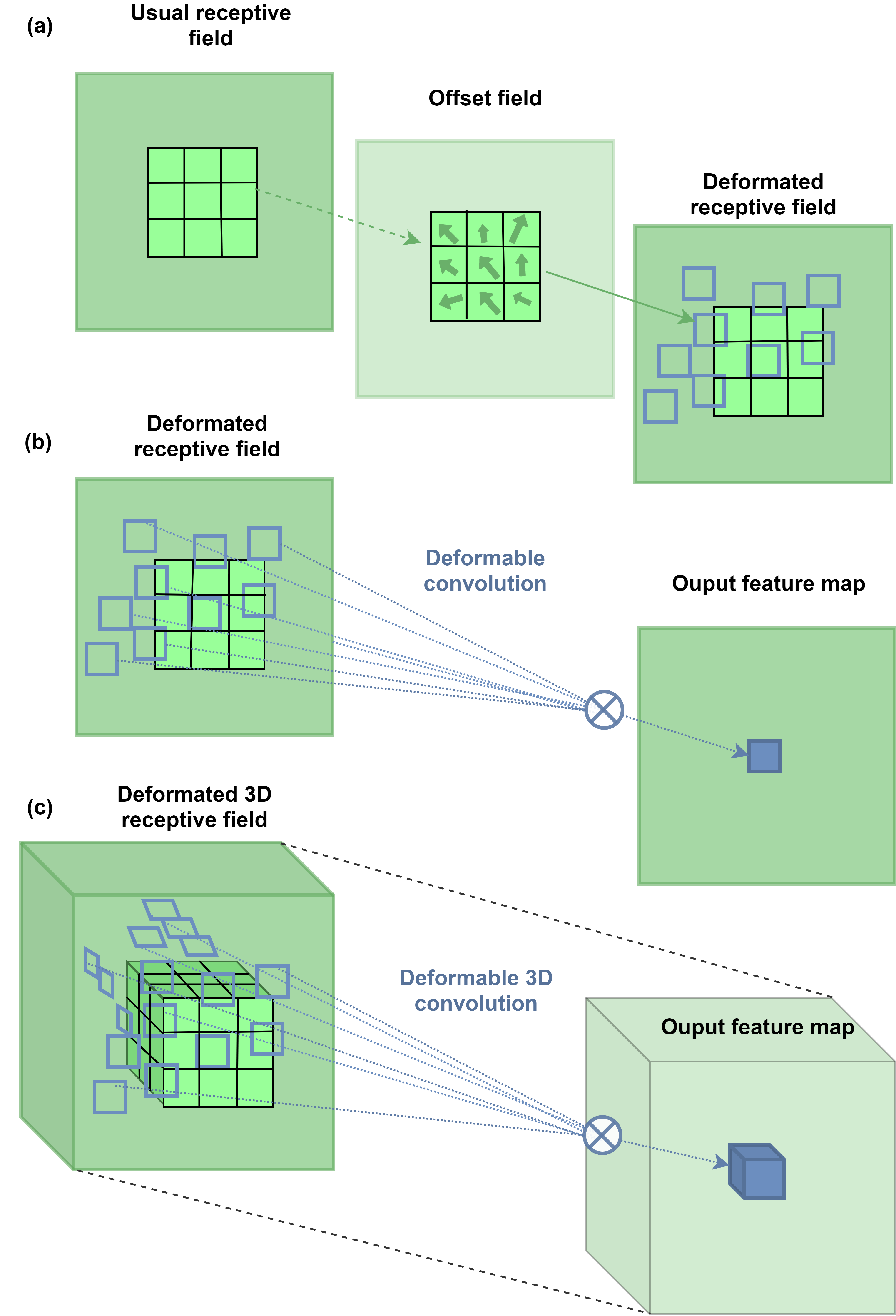}}
\caption{Illustration of deformable convolution: (a) 2D d-convolution [3, 3], the arrow in the offset field corresponds to how the blue squares are shifted in the input feature map; (b) 2D d-convolution; (c) 3D d-convolution [3, 3, 3].}
\label{fig:3d_dcov}
\end{center}
\vskip -0.2in
\end{figure}

The key difference between regular convolution and its deformable counterpart is the ability to deform the sampling grid for the standard convolution by predicting offsets for sampling locations. The offsets are learned from the activations of preceeding layer using additional convolutional layer, separately for each point of the activation map where the kernel is applied. The Fig.~\ref{fig:3d_dcov} illustrates how the offsets for the deformation are learned from the previous layer activation maps in the case of two- and three-dimensional image.

% an additional layer that is used to predict sampling locations. This layer is typically another convolution that outputs coordinates of offsets for each sampling location at each point where the filter is applied.

% the offsets are obtained by applying a convolutional layer over the same input feature map. The convolution kernel is of the same spatial resolution and dilation as those of the current convolutional layer. The output offset fields have the same spatial resolution with the input feature map. The channel dimension 2N corresponds to N 2D offsets. During training, both the convolutional kernels for generating the output features and the offsets are learned simultaneously. To learn the offsets, the gradients are backpropagated through the bilinear operations

The whole process for deformable convolution looks as follows. 
% Suppose, $R = \{(-1, -1), (-1, 0), \dots, (0, 1), (1, 1)\}$ is a sampling grid for the regular convolution (with kernel $3x3$ and dilation $1$ in this case). 
Suppose, $R$ is a sampling grid for the regular convolution. For each location $p_n$ on the input feature map $x$ and location $p_0$ in the sampling grid, an additional convolutional layer, typically with kernel of the same size and dilation, predicts an offset $\Delta p_n, \ n = 1, \dots, |R|$. Then the sampling grid for deformable convolution is augmented with predicted offsets, and kernel is applied to the values at locations $p_n + \Delta p_n$, which are computed via bilinear interpolation, since the obtained locations can be fractional. The final value, computed by deformable convolution at a point $p_0$ is $\sum\limits_{p_n \in R} w_n x(p_0 + p_n + \Delta p_n)$, where $w_n, n = 1, \dots, |R|$ are weights of the convolution kernel, instead of $\sum\limits_{p_n \in R} w_n x(p_0 + p_n)$ in the case of standard one.

\begin{figure}[ht]
\vskip 0.2in
\begin{center}
\centerline{\includegraphics[width=\columnwidth]{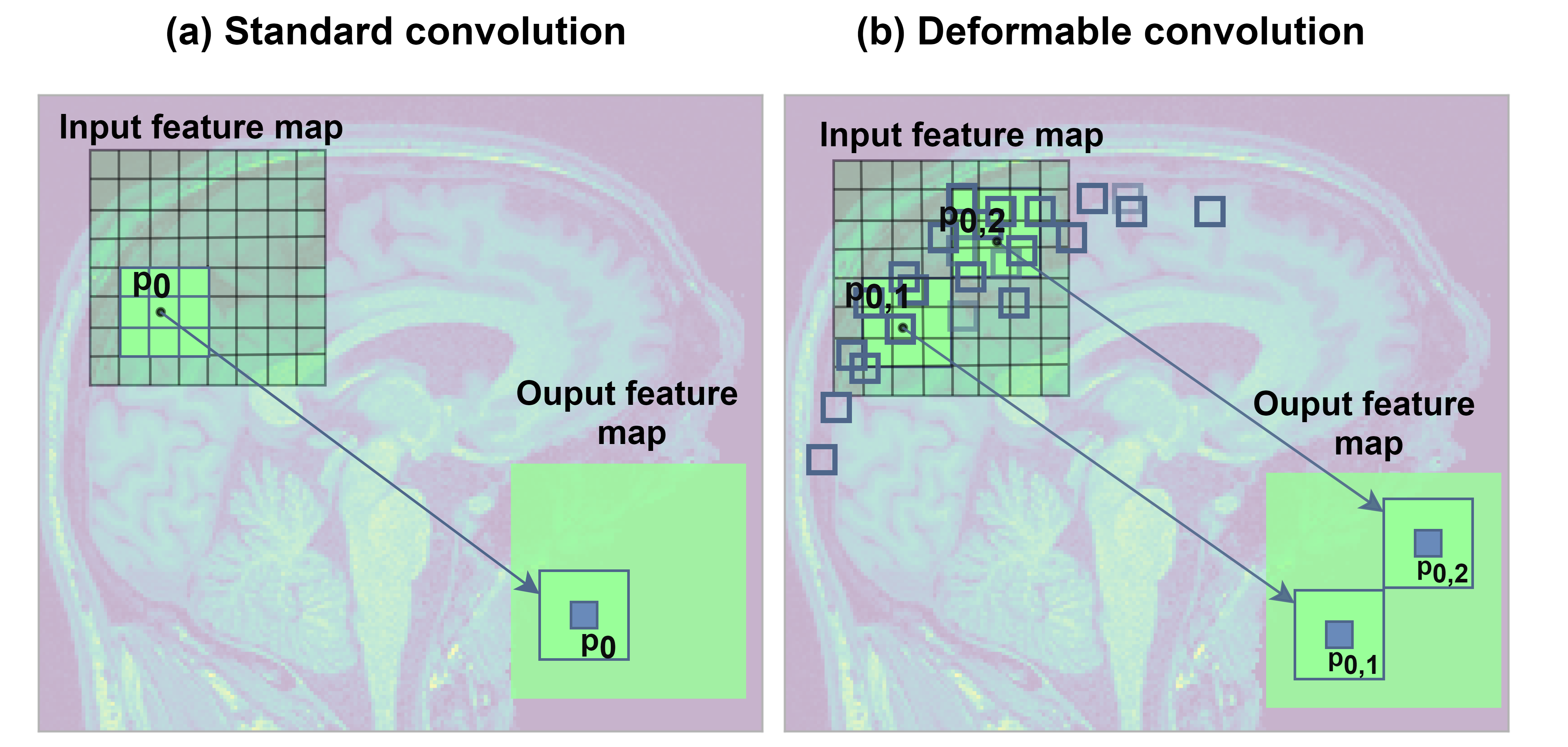}}
\caption{ Standard (a) and d-convolutions (b) applied to unprocessed brain imagery (Sagittal section). Grid represents pixel distribution within a sample in 2D case: (a) kernel with static receptive field; (b) d-convolution offsets with dynamic and learnable receptive field.}
\label{fig:def_brains}
\end{center}
\vskip -0.2in
\end{figure}

As a result, the receptive field of the convolutional kernel can change accordingly to the deformation of the input feature map and thus adapt to the variations of the size and scales of the learned distinguishing patterns for classification. On the Fig.~\ref{fig:def_brains} we can observe how the receptive field of deformable convolution differs from the regular one due to its adaptive sampling grid in case of brain imagery.

% \begin{figure}[ht]
% \vskip 0.2in
% \begin{center}
% \centerline{\includegraphics[width=\columnwidth]{3d_dconv.png}}
% \caption{Illustration of deformable convolution: (a) 2D d-convolution [3, 3], the arrow in the offset field corresponds to how the blue squares are shifted in the input feature map; (b) 2D d-convolution; (c) 3D d-convolution [3, 3, 3].}
% \label{fig:3d_dcov}
% \end{center}
% \vskip -0.2in
% \end{figure}

\subsection{dVoxResNet architecture}

\begin{figure}[ht]
\vskip 0.2in
\begin{center}
\centerline{\includegraphics[width=\columnwidth]{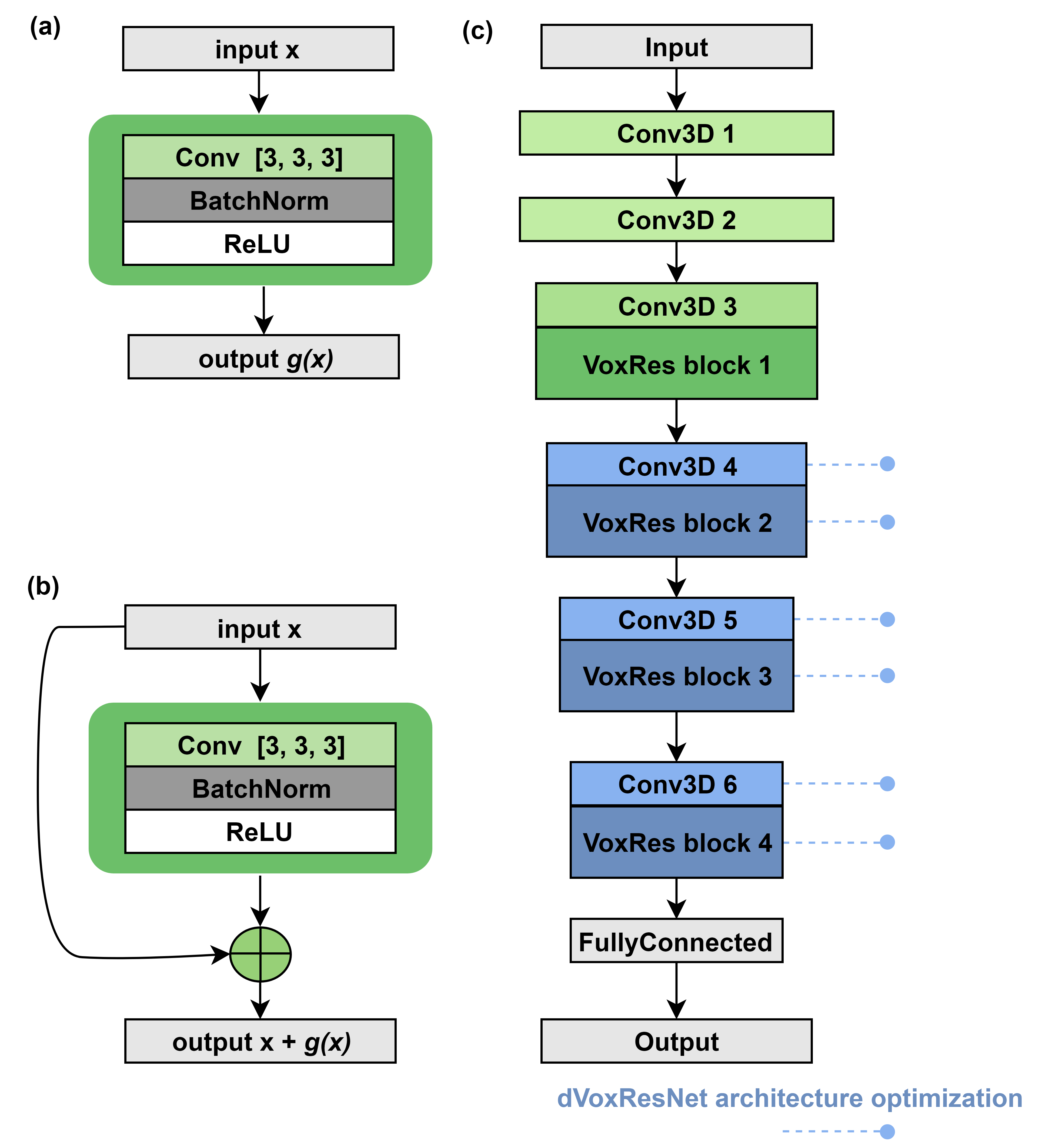}}
\caption{Architecture of dVoxResNet model used for structural MRI data classification: (a) \textit{Conv3D} unit structure; (b) \textit{VoxRes} unit structure; (c) \textit{dVoxResNet} model architecture, blue nodes represent chancing blocks for ablation study of design choices used for model optimisation.}
\label{fig:voxcnn}
\end{center}
\vskip -0.2in
\end{figure}

% For classification of schizophrenia and bipolar disorder versus healthy control
For the structural MRI data classification, we applied a modification of VoxResNet architecture consisting of 6 convolutional layers and 4 VoxRes blocks with two convolutional layers in each. First, we obtained baseline classification performances using standard model with only regular convolutional kernels. Next, we replaced regular three-dimensional  convolutions in one or several layers of the network with their deformable counterparts.
The architecture of our VoxResNet model is shown on Fig. \ref{fig:voxcnn}. We tested inserting deformable convolutions in the layers Conv3D from 4 to 6 and in both convolutional layers of VoxRes blocks from 2 to 4. Moreover, to study the effect of stacked deformable convolutions, we tried to apply them in several sequential layers and blocks of the network.

Due to the small sample sizes (172 subjects) we compare the classification results significance with paired non-parametric \textit{ttest} on ROC/AUC scores for repeated 5-fold cross validation (see description of a general pipeline in \cite{Pipeline2018}). In~\cite{beleites2005variance} it was shown, that stratified 5-fold CV had remarkably low bias compared to CV without repeats. Thus the variance could be reduced by repeating the n-fold error estimation over more than one random split of the data.

% We aimed at comparing 3D CNN with traditional convolution with a deformable one, for this purpose we adapt the baseline 3DCNN architecture by replacing regular filters in the middle layers with three-dimensional deformable convolutions. Then we trained it on augmented with random affine transformations data from the main domain and next applied the obtained network to classify data from another domain.

The models were implemented in \texttt{PyTorch} and trained on a single GPU \cite{canziani2016analysis}.

\subsection{Data and preprocessing}

\textit{dVoxResNet} performance was explored on a classification task between patients with Schizophrenia and Bipolar Disorder (as the most widespread psychiatric disorders) and healthy controls based on brain structural images. We aimed at finding a relatively big public dataset with multiple disorders collected at the same site.

\textbf{Dataset:} The data was collected from open databases from \textit{OpenNeuro (OpenfMRI)}\footnote{https://openneuro.org/, https://www.openfmri.org/} platform. 

The \textit{Main} dataset is retrieved from \textit{Consortium for Neuropsychiatric Phenomics study}\footnote{https://www.openfmri.org/dataset/ds000030/} \cite{gorgolewski2017preprocessed} for two pathologies: Schizophrenia and Bipolar Disorder as one of most spread psychiatric disorders which are known to have corresponding bio-markers or pathology patterns in brain structure \cite{zeng2012identifying, dakka2017learning}.

The additional dataset \textit{Working memory in healthy and schizophrenic individuals} \footnote{https://www.openfmri.org/dataset/ds000115/} \cite{repovs2012working} was used as a reference dataset for \textit{dVoxResNet} performance check.

We consider three classification problems: two main binary classifications from one source and one additional test dataset.

\begin{itemize}
    \item \textit{Dataset 1 (Main)}: 50 Schizophrenia patients vs 122 Healthy Controls;
    \item \textit{Dataset 2 (Main)}: 49 Bipolar Disorder patients vs 122 Healthy Controls;
    \item \textit{Dataset 3 (Additional)}: 47 Schizophrenia vs 34 Healthy Controls.
\end{itemize}

For each dataset the goal was to predict whether a subject is from pathology group or healthy control group.

\textbf{Preprocessing:} We performed different preprocessing steps included in regular structural MRI data preprocessing pipeline:

\begin{itemize}
    \item data anonymization;
    \item skull-stripping;
    \item skull-stripped brain normalization.
\end{itemize}

First, \texttt{Pydeface}\footnote{https://github.com/poldracklab/pydeface} was applied to all anatomical images to ensure deindentification of subjects. No additional preprocessing was applied on this stage, see Fig.\ref{fig:brain_preprocessing} (a). 

Second, defaced data was scull stripped with \texttt{FSL/BET}  \footnote{https://fsl.fmrib.ox.ac.uk/fslcourse/lectures/practicals/intro2/index.html} brain extraction toolbox   \cite{jenkinson2005bet2}. Thus the data from the second stage had no potentially uninformative features, see Fig.\ref{fig:brain_preprocessing} (b). 

On the third step brain images were normalized to MNI space \cite{laitinen1989co}, which implies standardization and alignment to common space for brain volumetric analysis during standard preprocessing protocol \textit{FreeSurfer-derived segmentations of the cortical gray-matter of Mindboggle} \cite{klein2017mindboggling} in \textit{fmriprep} \footnote{https://github.com/poldracklab/fmriprep} \cite{esteban2019fmriprep} toolbox. The result of the procedure can be seen at Fig.\ref{fig:brain_preprocessing} (c). 

After normalization, MR images were segmented in order to extract structural morphometric features (brain volumes, thicknesses, curvatures, number of vertexes, average voxel intensities within the region, etc.) which together form a feature vector. 

Structural features are calculated from $T1$ images using default processing pipeline in Freesurfer~\cite{fischl2012freesurfer} toolbox. Thus, morphometrical characterinstics (volumes, surface areas, thicknesses, etc.) are calculated for $34$ cortical regions according to Desikan-Killiany Atlas and for $45$ subcortical areas according to Automatic subcortical segmentation ~\cite{fischl2002whole} resulting in~a~vector of $927$ features for each subject.

The \textit{\textbf{Baseline performance}} was calculated with classification pipeline described in \cite{sharaev2018mri} with hyper-parameter grid search across classifiers: Logistic Regression, Support Vector Machine, Random Forest Classifier, k-Nearest Neighbours Classifier. The best model was \texttt{SVC, C = 10, kernel = rbf, gamma = 0.01}  implemented in \texttt{Sklearn}.

In order to test the dependence of classification performance on the level of images preprocessing, we ran classifiers on data after each preprocessing step, i.e. no preprocessing, skull-stripping, skull-stripping with normalization to MNI space.

\section{Results}

For both classification problems: Schizoprenia (50) vs Control (122) and Bipolar Disorder (49) vs Control (122) deep neural network classification performance is lower on MNI-normalized images. 

For Schizophrenia classification all deep neural methods outperform the Baseline classifier on morphometry features, see \textsc{Table I} where baseline classification AUC is \textbf{0.739 (0.086)}. All considered convolutional network architectures outperform the baseline on imagery without normalization. Optimized \textit{dVoxResNet} shows statistically significant performance increase according to the cross-validation scores.

For Bipolar Disorder classification deep neural methods show no significant difference from baseline classifier performance  of \textbf{0.668 (0.074)}, see \textsc{Table II}. Yet the optimized architecture of \textit{dVoxResNet} shows statistically significant performance increase yielding \textbf{0.687 (0.065)} on skull-stripped brain volumes.

The explicit description on model optimisation and corresponding cross validation scores for dVoxResNet model performance depending architecture studied on schizophrenia and Bipolar Disorder classification are shown in \textsc{Table III}, see \textsc{Appendix}.

\begin{center}
    
\begin{table}[t!]

\caption{Schizophrenia classification: Schizophrenia (50) / Control (122). VoxResNet model with and without d-convolutional layers on different prepossessing stages. \textit{ Validated on 3-fold CV with 3 repeats, ROC/AUC}.}
\begin{tabular}{l|lll}
\toprule
\textit{\begin{tabular}[c]{@{}l@{}}MRI data type \\ and preprocessing.\\  \end{tabular}} & \begin{tabular}[c]{@{}l@{}}3D CNN,\\ Mean (STD)\end{tabular} & \begin{tabular}[c]{@{}l@{}} dVoxResNet, \\ Mean (STD)\end{tabular} & \begin{tabular}[c]{@{}l@{}} dVoxResNet \\ optimised, \\ Mean (STD)\end{tabular} \\
\hline
\\
No prepossessing & 0.788 (0.068) & 0.808 (0.064) & {\color[HTML]{009901} \textbf{0.823 (0.058)}} \\
skull-stripping & 0.778 (0.075) & 0.806 (0.055) & {\color[HTML]{009901} \textbf{0.823 (0.065)}} \\
\begin{tabular}[c]{@{}l@{}}skull-stripping, \\MNI normalization\end{tabular} & 0.736  (0.070) & 0.737 (0.063) & 0.731 (0.063) \\
\\
\hline
\textit{\textbf{\begin{tabular}[c]{@{}l@{}}Baseline: \\morphometry features\end{tabular}}} & \multicolumn{3}{c}{0.739 (0.086)}
\end{tabular}
\label{table:1}
\end{table}

\end{center}

\begin{table}[t!]
\centering

\caption{Bipolar disorder classification: Bipolar Disorder (50) / Control (122). VoxResNet model with and without d-convolutional layers on different prepossessing stages.  \textit{ Validated on 5-fold CV with 3 repeats, ROC/AUC} .}
\begin{tabular}{l|lll}
\toprule
\textit{\begin{tabular}[c]{@{}l@{}}MRI data type \\ and preprocessing.\\  \end{tabular}} & \begin{tabular}[c]{@{}l@{}}3D CNN,\\ Mean (STD)\end{tabular} & \begin{tabular}[c]{@{}l@{}} dVoxResNet, \\ Mean (STD)\end{tabular} & \begin{tabular}[c]{@{}l@{}}dVoxResNet \\ optimised, \\ Mean (STD)\end{tabular} \\
\hline
\\
No prepossessing & 0.639 (0.088) & 0.639 (0.088) & {\color[HTML]{009901} \textbf{0.676 (0.081)}} \\
skull-stripping & 0.648 (0.102) & 0.651 (0.065) & {\color[HTML]{009901} \textbf{0.687 (0.065)}} \\
\begin{tabular}[c]{@{}l@{}}skull-stripping, \\ MNI normalization\end{tabular} & 0.639 (0.105) & 0.629 (0.065) & 0.631 (0.097) \\
\\
\hline
\textit{\textbf{\begin{tabular}[c]{@{}l@{}}Baseline:  \\morphometry features\end{tabular}}} & \multicolumn{3}{c}{0.668 (0.074)}
% \bottomrule
\end{tabular}
\label{table:2}
\end{table}

\section{Discussion}

It can be seen that insertion  of  deformable  convolution  layers instead of traditional ones in  VoxResNet architecture yield statistically significant improvement in classification  accuracy  for  both schizophrenia and bipolar disorder classification tasks.  Moreover, d-convolutions perform well on MRI data without  preprocessing  and skull-stripping.
Our experiments with \textit{dVoxResNet} architecture also revealed decrease of classification accuracy of schizophrenia and bipolar disorder with increaing the level of preprocessing. This could potentially arise from the fact that data cleaning and preprocessing removes informative parts of data. Given that typical preprocessing pipeline for structural imagery is computationally expensive and takes up to 15 min per subject for skull-stripping and up to 5 hours per subject for normalization in FreeSurfer (on a single CPU core), applying d-convolutions seems beneficial and the need for thourough MR images preprocessing for classification with convolutional networks should be investigated. 

Insertion of deformable convolution layers in VoxResNet architecture yield statistically significant improvement in classification accuracy for imagery without preprocessing and skull-stripping of images.
However, they do not improve classification performance on preprocessed data. This can be possibly explained by the fact that MNI normalization significantly reduces variability in small deformations of the brain image and deformable convolutions lack information for training the meaningful offset. But at the same time the number of trainable parameters increases, making the model with deformable convolutions less stable to overfitting.

In case of disease recognition based on unprocessed or skull-stripped data, the use of deformable convolutions in only one layer or block gives only a slight quality improvement, and thus is potentially not enough to provide the network with the necessary deformation modelling capability. However, the use of stacked deformable convolutions in several sequential layers already allows to obtain the effect of a statistically significant improvement of classification performance. These results are also consistent with the observations for deformation modelling with 2D deformable ConvNets \cite{zhu2019deformable}.

We also applied deformable ConvNets for schizophrenia recognition problem on \textit{Dataset 3 (Additional)} of smaller size, see \textsc{Table IV} in \textsc{Appendix}. We tested both training the model from scratch on this small sample and transferring and fine-tuning the pre-trained weights for Schizophrenia versus Healthy control classification from the main dataset. Yet it does not show any significant classification improvement, which can be due to small sample size causing in poor model generalizability.

\subsection{Limitations}

The current study has several limitations mostly resulting from the computational cost of deformable convolutions and 3D networks on MRI volumes. The d-convolutions are to be used with the data augmentation yet it more computationally expensive.

The use of convolutions with bigger than [3, 3, 3] kernels also was restricted due to \textit{GPU} memory capacity.

Limited data: 122 control, 50 Schizophrenia and 49 Bipolar Disorder subjects; larger samples classification will allow more disperse estimation of model performance, without computational expenses on cross validation. 

\subsection{Further directions}

% An improvement of classification performance probably could be reached with data augmentation for training. As d-convolutions are potentially insensitive to image deformations, 

Deformable convolution tries to learn how to predict the sampling locations in a way to make filters invariant to small deformations. Thus, augmentation of brain imagery data using small scales and affine transformations can add more variability and provide additional information for training networks with deformable convolutional modules. It could also be potentially  helpful to get more stable results by increasing the training sample size. However, applying such augmentation is more computationally expensive.

Also wide learnable receptive field could be beneficial on first layers, kernels with size [7, 7, 7] or [11, 11, 11] can be more effective for global patterns then [3, 3, 3] kernels. 

% Yet they require extended \textit{GPU} memory capacity: for each feature in kernel the offset map should be calculated in three directions to define deformed receptive field in volume, which requires massive computations. 
The use of deformable convolutions with bigger than $3\times 3\times3$ kernels requires extended \textit{GPU} memory capacity. Prediction of the offsets for each unit of the convolutional kernel of size $k$ at each point of the input feature map assumes generation of the additional $3 \times k^3$ activation maps.

D-convolutions now explored for MRI classification problems were originally proposed for image segmentation tasks, and can be further utilized for brain segmentation as well.

Apart from deformable convolutions, transformable convolutions \cite{xiao2018transformable} can improve deep neural model performance. Besides dynamic sampling matrix in transformable convolutions there is a static global sampling matrix and they are used together for getting the output feature map. Thus defining global offset map depending on a brain structure we can introduce a domain specific and improve accuracy. Other lines of research could include usage of sparse convolutions \cite{3DCNN2018} for computational efficiency and fusion of multi-fidelity data \cite{Multispectral2018} to increase prediction accuracy.

The additional study would be a great of interest both for potential accuracy improvement and neural networks results interpretation.

\section{Conclusion}

We proposed new 3D deformable convolutions (d-convolutions) application for structural MRI classification task and implemented it in VoxResNet architecture --- \textit{dVoxResNet}.

The usage of deformable convolution layers yields statistically significant improvement in  classification performance for unprocessed and skull-stripped brain images. Yet it was not effective for normalized brain images, which could potentially arise from  the  fact  that  data  normalization  removes
informative  parts  of  data and reduces variability. 
We showed that deformable convolutions could be competitive analogues of standard ones, and their usage is reasonable despite of the high computational cost. 3D d-convolutions significantly outperform standard ones in binary classification tasks and are effective for unprocessed 3D images. 

Firstly proposed \textit{dVoxResNet} architecture show a high potential for application to other psycho-neurological disorders diagnostics from different datasets.

\section*{Acknowledgment}
The study was supported by the Russian Science Foundation under Grant 19-41-04109.

%This study was performed in the scope of the Project “Machine Learning and Pattern Recognition for the development of diagnostic and clinical prognostic prediction tools in psychiatry, borderline mental disorders, and neurology” (a part of the Skoltech Biomedical Initiative program).

\bibliography{references.bib}
\bibliographystyle{ieeetr}

\clearpage
\onecolumn

\appendix

\begin{table*}[ht!]
\centering
\caption{dVoxResNet model optimization: model performance depending architecture studied on schizophrenia and Bipolar Disorder classification \textit{ Validated on 3-fold CV, ROC/AUC}. d-convolutions are inserted in blocks according to \textit{idx}: Conv3D blocks [1:6], VoxRes blocks [1:4], where colour defines \color[HTML]{009901} \textbf{statistically significant performance increase},  \color[HTML]{ee7777} \textbf{statistically significant performance decrease} }
\begin{tabular}{@{}c|cccccc@{}}
\toprule
 & \multicolumn{3}{c}{Sсhizophrenia (50)/ Healthy Control (122)} & \multicolumn{3}{c}{Bipolar Disorder (49)/ Healthy Control (122)} \\
\begin{tabular}[c]{@{}c@{}} Conv3D block idx; \\ VoxRes block idx\end{tabular} & unprocessed & skull - striped & MNI normalized & unprocessed & skull - striped & MNI normalized \\
\hline
Image input size & 176x200x152 & 176x200x152 & 144x176x144 & 176x200x152 & 176x200x152 & 144x176x144 \\
\hline
\textbf{- ; -} & \textbf{0.788 +/- 0.068} & \textbf{0.778 +/- 0.075} & \textbf{0.736 +/- 0.070} & \textbf{0.639 +/- 0.088} & \textbf{0.648 +/- 0.102} & \textbf{0.639 +/- 0.105} \\
4 ; - & {\color[HTML]{000000} 0.808 +/- 0.064} & 0.806 +/- 0.055 & {\color[HTML]{000000} 0.737 +/- 0.063} & {0.675 +/- 0.096} & 0.651 +/- 0.065 & 0.629 +/- 0.065 \\
- ; 2 & 0.790 +/- 0.070 & 0.797 +/- 0.064 & {\color[HTML]{000000} 0.734 +/- 0.091} & {\color[HTML]{000000} 0.654 +/- 0.084} & 0.654 +/- 0.104 & 0.658+/- 0.099 \\
5 ; - & 0.797 +/- 0.076 & 0.805 +/- 0.065 & {\color[HTML]{000000} 0.736 +/- 0.082} & {\color[HTML]{000000} 0.667 +/- 0.093} & {\color[HTML]{000000} 0.662 +/- 0.089} & 0.599 +/- 0.104 \\
- ; 3 & 0.801 +/- 0.074 & 0.802 +/- 0.066 & {\color[HTML]{000000} 0.721 +/- 0.071} & {0.680 +/- 0.093} & 0.637 +/- 0.081 & 0.629 +/- 0.073 \\
6; - & 0.783 +/- 0.072 & 0.802 +/- 0.068 & {\color[HTML]{000000} 0.739 +/- 0.062} & {\color[HTML]{000000} 0.638 +/- 0.107} & {0.674 +/- 0.092} & 0.622 +/- 0.088 \\
- ; 4 & 0.782 +/- 0.069 & 0.798 +/- 0.052 & {\color[HTML]{000000} 0.748 +/- 0.063} & {\color[HTML]{000000} 0.640 +/- 0.089} & {0.669 +/- 0.085} & 0.611 +/- 0.116 \\
4 ; 2 & {\color[HTML]{009901} \textbf{0.822 +/- 0.068}} & 0.801 +/- 0.071 & 0.734 +/- 0.071 & {0.671 +/- 0.095} & {0.668 +/- 0.110} & {\color[HTML]{ee7777} \textbf{0.594 +/- 0.075}} \\
4, 5 ; 2 & {\color[HTML]{009901} \textbf{0.817 +/- 0.045}} & 0.802 +/- 0.059 & 0.716 +/- 0.086 & {0.672 +/- 0.081} & 0.638 +/- 0.078 & 0.604 +/- 0.100 \\
4, 5 ; 2, 3 & {\color[HTML]{009901} \textbf{0.823 +/- 0.058}} & {\color[HTML]{009901} \textbf{0.823 +/- 0.065}} & 0.731 +/- 0.063 & {0.679 +/- 0.062} & 0.638 +/- 0.088 & {\color[HTML]{ee7777} \textbf{0.571 +/- 0.091}} \\
5 ; 2, 3 & {\color[HTML]{009901} \textbf{0.815 +/- 0.072}} & 0.803 +/- 0.054 & 0.745 +/- 0.078 & {\color[HTML]{000000} 0.669 +/- 0.092} & 0.655 +/- 0.079 & 0.659 +/- 0.092 \\
5 ; 3 & {\color[HTML]{000000} 0.805 +/- 0.077} & {\color[HTML]{009901} \textbf{0.812 +/- 0.062}} & {\color[HTML]{000000} 0.720 +/- 0.089} & {\color[HTML]{009901} \textbf{0.681 +/- 0.075}} & {\color[HTML]{000000} 0.659 +/- 0.069} & 0.659 +/- 0.092 \\
5, 6 ; 3 & 0.793 +/- 0.067 & 0.807 +/- 0.066 & {\color[HTML]{000000} 0.727 +/- 0.087} & {0.676 +/- 0.064} & {\color[HTML]{000000} 0.661 +/- 0.083} & 0.626 +/- 0.091 \\
5, 6 ; 3, 4 & 0.797 +/- 0.072 & {\color[HTML]{009901} \textbf{0.816 +/- 0.060}} & 0.722 +/- 0.072 & {\color[HTML]{009901} \textbf{0.676 +/- 0.081}} & {\color[HTML]{009901} \textbf{0.687 +/- 0.065}} & 0.631 +/- 0.097 \\
6 ; 3, 4 & 0.795 +/- 0.079 & 0.805 +/- 0.071 & 0.726 +/- 0.064 & {\color[HTML]{000000} 0.656 +/- 0.092} & {\color[HTML]{000000} 0.660 +/- 0.091} & 0.606 +/- 0.097 \\ 
\\
\end{tabular}
\label{table:3}
\end{table*}

% \end{multicols}
\begin{table}[ht!]

\caption{Schizophrenia classification on \textbf{Main} and \textbf{Additional} datasets. Comparison of transfer of pre-trained weights and fine-tuning the VoxResNet model for schizophrenia versus Control classification from Dataset 1 (Main) to smaller test Dataset 3 (Additional)}
\centering

\begin{tabular}{c|cccc}
\hline
                                & \multicolumn{4}{c}{5-fold CV with 3 repeats ROC AUC score, Mean (STD)}                                                    \\ \hline
Sсhizophrenia / Healthy Control &       \textbf{Main} unprocessed                     & \multicolumn{3}{c}{\textbf{Additional} unprocessed}                                    \\ \cline{1-1}
Conv3D block idx; VoxRes block idx   & dVoxResNet                                    & dVoxResNet               & Weights transfer from \textbf{Main} & Finetuning             \\ \hline
\textbf{- ; -}                  & \multicolumn{1}{c|}{\textbf{0.794 (0.068)}} & \textbf{0.742 (0.150)} & \textbf{0.669}            & \textbf{0.749 (0.128)} \\
4 ; 2                           & \multicolumn{1}{c|}{0.822 (0.068)}          & 0.764 (0.113)          & 0.681                     & 0.740 (0.124)          \\
4, 5 ; 2                        & \multicolumn{1}{c|}{0.817 (0.045)}          & 0.753 (0.143)          & 0.653                     & 0.755 (0.126)          \\
4, 5 ; 2, 3                     & \multicolumn{1}{c|}{0.823 (0.058)}          & 0.746 (0.114)          & 0.621                     & 0.676 (0.140)          \\
5 ; 2, 3                        & \multicolumn{1}{c|}{0.815 (0.072)}          & 0.712 (0.125)          & 0.646                     & 0.759 (0.113)          \\ 
\\
\end{tabular}
\label{table:4}
% \caption{Schizophrenia classification on \textbf{Main} and \textbf{Additional} datasets. Comparison of transfer of pre-trained weights and fine-tuning the VoxResNet model for schizophrenia versus Control classification from Dataset 1 (Main) to smaller test Dataset 3 (Additional)}
\end{table}

\end{document}